\documentclass[11pt,a4paper]{article}

\usepackage{inputenc}
\usepackage{amsmath}
\usepackage{bm}
\usepackage{bbold}
\usepackage{amsthm}

\usepackage[hyphens]{url}

\usepackage{hyperref}
\usepackage{breakurl}

\title{Modeling and Performance Evaluation\\ of Computer Systems Security Operation\thanks{Proc. 4th St.~Petersburg Workshop on Simulation / Ed. by S.~M.~Ermakov, Yu.~N.~Kashtanov, V.~B.~Melas, NII Chemistry St.~Petersburg University Publishers, St.~Petersburg, 2001, pp.~233--238.}} 

\author{D.~Guster\thanks{Department of Statistics, St.~Cloud State University, 720 4th Ave. S., St.Cloud, MN 56301-4442, Guster@mcs.stcloudstate.edu.}
\and
N.~K.~Krivulin\thanks{Faculty of Mathematics and Mechanics, St.~Petersburg State University, 28 Universitetsky Ave., St.~Petersburg, 198504, Russia, 
nkk@math.spbu.ru.} \thanks{The work was partially supported by the Russian Foundation for Basic Research, Grant~\#00-01-00760.}
}

\date{}

\newtheorem{theorem}{Theorem}
\newtheorem{lemma}[theorem]{Lemma}

\begin{document}

\maketitle

\begin{abstract}
A model of computer system security operation is developed based on the
fork-join queueing network formalism. We introduce a security operation 
performance measure, and show how it may be used to performance evaluation of
actual systems.
\\

\end{abstract}

\section{Introduction}
The explosive growth in computer systems and networks has increased the role 
of computer security within organizations \cite{Stallings1995Network}. In many cases, 
ineffective protection against computer security treats leads to considerable 
damage, and even can cause an organization to be paralized. Therefore, the
development of new models and methods of performance analysis of security 
systems seems to be very important.

In this paper, we propose a model of computer security operation, and 
introduce its related performance measure. It is shown how the model can be 
applied to performance evaluation of actual systems. Finally, a technique of 
security system performance analysis is described and its practical 
implementation is discussed.

We conclude with an appendix which contains technical details concerning 
fork-join network representation of the model, and related results.

\section{A Security Operation Model}

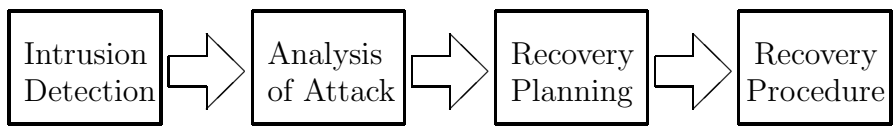
\begin{figure}[ht]
\begin{center}
%\vspace{-1ex}
\setlength{\unitlength}{1mm}
\begin{picture}(115,15)
\newsavebox\textbox
\savebox{\textbox}(20,15){\thicklines
 \put(0,0){\line(1,0){20}}
 \put(0,15){\line(1,0){20}}
 \put(0,0){\line(0,1){15}}
 \put(20,0){\line(0,1){15}}}

\newsavebox\bbarrow
\savebox{\bbarrow}(10,6){\thinlines
 \put(0,0){\line(1,0){5}}
 \put(0,4){\line(1,0){5}}
 \put(0,0){\line(0,1){4}}
 \put(5,4){\line(0,1){3}}
 \put(5,0){\line(0,-1){3}}
 \put(5,7){\line(1,-1){5}}
 \put(5,-3){\line(1,1){5}}}

\put(0,0){\usebox\textbox}

\put(2,8){\large Intrusion}
\put(2,3){\large Detection}

\put(21,4){\usebox\bbarrow}

\put(32,0){\usebox\textbox}

\put(34,8){\large Analysis}
\put(34,3){\large of Attack}

\put(53,4){\usebox\bbarrow}

\put(64,0){\usebox\textbox}

\put(66,8){\large Recovery}
\put(66,3){\large Planning}

\put(85,4){\usebox\bbarrow}

\put(96,0){\usebox\textbox}

\put(98,8){\large Recovery}
\put(97,3){\large Procedure}

\end{picture}

%\vspace{-2ex}
\end{center}
\caption{Computer systems security activities.}\label{F-CSSA}
\end{figure}

In this paper, we deal with the current security activities 
(see Fig.~\ref{F-CSSA}) that mainly relate to the actual security threats 
rather than to strategic or long-term issues of security management. 

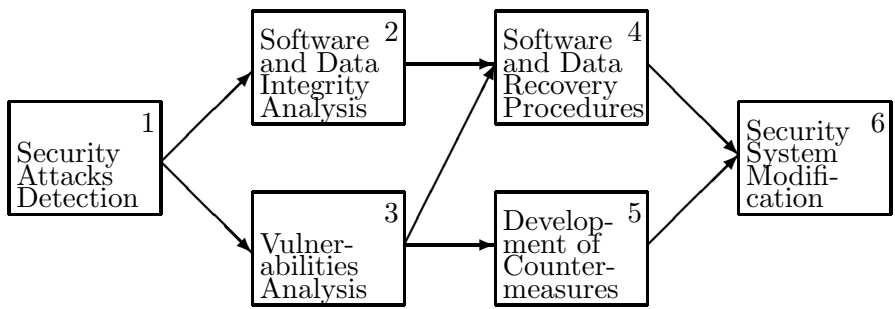
\begin{figure}[ht]
\begin{center}
%\vspace{-1ex}
\setlength{\unitlength}{1mm}
\begin{picture}(115,40)
\savebox{\textbox}(20,15){\thicklines
 \put(0,0){\line(1,0){20}}
 \put(0,15){\line(1,0){20}}
 \put(0,0){\line(0,1){15}}
 \put(20,0){\line(0,1){15}}}

\put(17,23){$1$}
\put(0,12){\usebox\textbox}

\put(1,19){Security}
\put(1,16){Attacks}
\put(1,13){Detection}

\put(20,19){\thicklines\vector(1,1){12}}
\put(20,19){\thicklines\vector(1,-1){12}}

\put(49,11){$3$}
\put(32,0){\usebox\textbox}

\put(33,7){Vulner-}
\put(33,4){abilities}
\put(33,1){Analysis}

\put(52,8){\thicklines\vector(1,2){12}}
\put(52,8){\thicklines\vector(1,0){12}}

\put(49,35){$2$}
\put(32,24){\usebox\textbox}

\put(33,34){Software}
\put(33,31){and Data}
\put(33,28){Integrity}
\put(33,25){Analysis}

\put(52,32){\thicklines\vector(1,0){12}}

\put(81,11){$5$}
\put(64,0){\usebox\textbox}

\put(65,10){Develop-}
\put(65,7){ment of}
\put(65,4){Counter-}
\put(65,1){measures}

\put(84,8){\thicklines\vector(1,1){12}}

\put(81,35){$4$}
\put(64,24){\usebox\textbox}

\put(65,34){Software}
\put(65,31){and Data}
\put(65,28){Recovery}
\put(65,25){Procedures}

\put(84,32){\thicklines\vector(1,-1){12}}

\put(113,23){$6$}
\put(96,12){\usebox\textbox}

\put(97,22){Security}
\put(97,19){System}
\put(97,16){Modifi-}
\put(97,13){cation}

\end{picture}

%\vspace{-2ex}
\end{center}
\caption{A security analysis and maintenance model.}\label{F-SAMM}
\end{figure}

Consider the model of security operation in an organization, presented in 
Fig.~\ref{F-SAMM}. Each operational cycle starts with security attack 
detection based on audit records and system/errors log analysis, traffic 
analysis, or user reports. In order to detect an intrusion, automated tools of 
security monitoring are normally used including procedures of statistical 
anomaly detection, rule-based detection, and data integrity control 
\cite{Stallings1995Network}.

After security attack detection and identification, the integrity of 
system/application software and data in storage devices has to be examined 
to search for possible unauthorized modifications or damages made by the 
intruder. The investigation procedure can exploit file lists and checksum 
analysis, hash functions, and other automated techniques.

In parallel, the system vulnerabilities, which allow the intruder to attack, 
should be identified and investigated. The vulnerability analysis normally 
presents an informal procedure, and therefore, it can hardly be performed 
automatically. 

Based on the results of integrity analysis, a software and data recovery 
procedure can be initiated using back-up servers and reserving storage 
devices. It has to take into account the security vulnerabilities identified 
at the previous step, so as to provide for further improvements in the entire 
security system.

Along with the recovery procedure, the development of a complete set of 
countermeasures against similar attacks should be performed. Finally, the 
operational cycle is concluded with appropriate modifications of software, 
data bases, and system security policies and procedures.

We assume that the organization has appropriate personnel integrated in a
Computer Emergency Response Team, available to handle the attack. The team 
would include at least two subteams working in parallel, one to perform 
integrity analysis and recovery procedures, and another to do vulnerability 
analysis and development of countermeasures. At any time instant, each 
subteam can deal with only one security incident. Any procedure may be started 
as soon as all prior procedures according to the model in Fig.~\ref{F-SAMM}, 
have been completed. If a request to handle a new incident occurs when a 
subteam is still working on a procedure, the request has to wait until 
the processing of that procedure is completed.

We denote by $ \tau_{1k} $ a random variable (r.v.) that represents the
time interval between detections of the $k$th attack and its predecessor.
Furthermore, we introduce r.v.'s $ \tau_{ik} $, $ i=2,\ldots,6 $, to 
describe the time of the $k$th instant of procedure $ i $ in the model.
We assume $ \tau_{i1},\tau_{i2},\ldots $, to be independent and identically 
distributed (i.i.d.) r.v.'s with finite mean and variance for each $ i $,
$ i=1,\ldots,6 $. At the same time, we do not require of independence of 
$ \tau_{1k},\ldots,\tau_{6k} $ for each $ k $, $ k=1,2,\ldots $.

\section{Security Operation Performance Evaluation}

In order to describe system performance, we introduce the following notations.
Let $ \overline{T}_{A} $ be the mean time between consecutive security
attacks (the attack cycle time), and $ \overline{T}_{S} $ be the mean time
required to completely handle an attack (the recovery cycle time), as the 
number of attacks $ k $ tends to $ \infty $.

In devising the security operation performance measure, one can take the ratio
$$
R=\overline{T}_{S}/\overline{T}_{A}.
$$
With the natural condition $ \overline{T}_{S}\leq\overline{T}_{A} $, one can 
consider $ R $ as the time portion the system is under recovery, assuming
$ k\to\infty $.

First note that the attack cycle time can immediately be evaluated as the mean
value: $ \overline{T}_{A}={\rm E}[\tau_{11}] $. 

Now consider the cycle time of the entire system, which can be defined as
the mean time interval between successive completions of security system 
modification procedures as the number of attacks $ k\to\infty $. As one
can prove (see Appendix for further details), the system cycle time 
$ \gamma $ can be calculated as
$$
\gamma=\max\{{\rm E}[\tau_{11}],\ldots,{\rm E}[\tau_{61}]\}.
$$

In order to evaluate the recovery cycle time, we assume the system will 
operate under the maximum traffic level, which can be achieved when all the 
time intervals between attacks are set to $ 0 $. Clearly, under that 
condition, the system cycle time can be taken as a reasonable estimate of the 
recovery cycle time.

Considering that now $ {\rm E}[\tau_{11}]=0 $, we get the recovery cycle 
time in the form
$$
\overline{T}_{S}=\max\{{\rm E}[\tau_{21}],\ldots,{\rm E}[\tau_{61}]\}.
$$

\section{Performance Analysis and Discussion}

In fact, the above model presents a quite simple but useful tool for security
system operation management. It may be used to make decision on the basis of
a few natural parameters of the security operation process.

Let us represent the ratio $ R $ in the form
$$
R=\max\{{\rm E}[\tau_{21}],\ldots,{\rm E}[\tau_{61}]\}/{\rm E}[\tau_{11}],
$$
and assume the attack rate determined by $ {\rm E}[\tau_{11}] $, to be 
fixed. 

Taking into account that the above result has been obtained based on the
assumption of an infinite number of attacks, we arrive at the following 
conclusion. As the number of attacks becomes sufficiently large, the 
performance of the system is determined by the time of the longest procedure 
involved in the system operation, whereas the impact of the order of 
performing the procedures disappears. 

It is clear that in order to improve system performance, the system security 
manager (administrator) should first concentrate on decreasing the mean time 
required to perform the longest procedure within the security operation model, 
then consider the second longest procedure, and so on. The goal of decreasing 
the time can be achieved through partition of a whole procedure into 
subprocedures, which can be performed in parallel, or through rescheduling of 
the entire process with redistribution of particular activities between 
procedures.

In practice, the above model and its related ratio $ R $ can serve as the 
basis for efficient monitorization of organizational security systems. Because
the introduction of new countermeasures may change the attack cycle time, the 
monitoring requires updating this parameter after each modification of the 
system.

Finally note, the above model can be easily extended to cover security 
operational processes, which consist of different procedures and precedence 
constraints.

\section*{Appendix}
In order to describe the above security system operational model in a formal 
way, we exploit the fork-join network formalism proposed in \cite{Baccelli1989Queueing}.

The fork-join networks present a class of queueing systems, which allow for
splitting a customer into several new customers at one node, and of merging 
customers into one at another node. In order to represent the dynamics of such
networks, we use a $(\max,+)$-algebra based approach developed in 
\cite{Krivulin2000Algebraic}. 

The $(\max,+)$-algebra is a triple 
$ \langle R_{\varepsilon},\oplus,\otimes\rangle $, where
$ R_{\varepsilon}=R \cup \{\varepsilon\} $ with
$ \varepsilon=-\infty $. The operations $ \oplus $ and $ \otimes $
are defined for all $ x,y\in R_{\varepsilon} $ as
$$
x \oplus y=\max(x,y), \quad x \otimes y=x+y.
$$

The $(\max,+)$-algebra of matrices is introduced in the ordinary way with the
matrix $ {\cal E} $ with all its entries equal $ \varepsilon $, taken as 
the null matrix, and the matrix $ E={\rm diag}(0,\ldots,0) $ with its 
off-diagonal entries equal $ \varepsilon $, as the identity.

We introduce the vector 
$ \mbox{\boldmath $x$}(k)=(x_{1}(k),\ldots,x_{n}(k))^{T} $ as the $k$th 
service completion times at the network nodes, and the diagonal matrix 
$ {\cal T}_{k}={\rm diag}(\tau_{1k},\ldots,\tau_{nk}) $ with given 
nonnegative random variables $ \tau_{ik} $ representing the $k$th service 
time at node $ i $, $ i=1,\ldots,n $, and the off-diagonal entries equal 
$ \varepsilon $.

The dynamics of acyclic fork-join networks can be described by the stochastic 
difference equation (see \cite{Krivulin2000Algebraic} for further details)
\begin{equation}\label{E-xAx}
\mbox{\boldmath $x$}(k)=A(k)\otimes\mbox{\boldmath $x$}(k-1),
\qquad
A(k)
=
\bigoplus_{j=0}^{p}({\cal T}_{k}\otimes G^{T})^{j}\otimes{\cal T}_{k},
\end{equation}
where $ G $ is a matrix with the elements
$$
g_{ij}=\left\{\begin{array}{ll}
        0, & \mbox{if there exists arc $ (i,j) $ in the network graph}, \\
        \varepsilon, & \mbox{otherwise},
       \end{array}\right.
$$
and $ p $ is the length of the longest path in the graph. 

The matrix $ G $ is normally referred to as the support matrix of the
network. Note that since the network graph is acyclic, we have 
$ G^{q}={\cal E} $ for all $ q>p $.

The cycle time of the network is defined as 
$$
\gamma=\lim_{k\to\infty}\|\mbox{\boldmath $x$}(k)\|^{1/k},
$$
where $ \|\mbox{\boldmath $x$}(k)\|=\max_{i}x_{i}(k) $. Clearly, if this 
limit exists, it can be found as $ \lim_{k\to\infty}\|A_{k}\|^{1/k} $, where 
$ A_{k}=A(k)\otimes\cdots\otimes A(1) $.

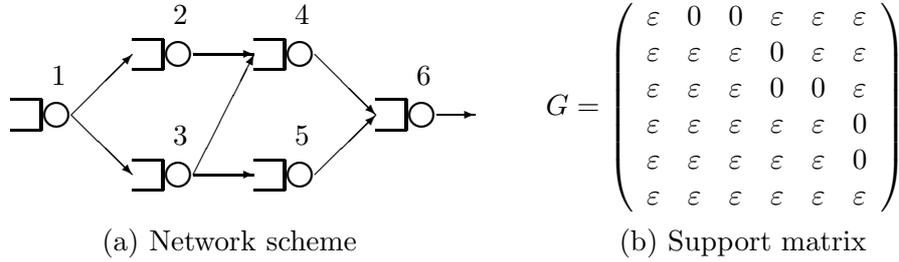
\begin{figure}[ht]
\begin{center}
%\vspace{-1ex}
\setlength{\unitlength}{1mm}
\begin{picture}(115,35)
\newsavebox\queue
\savebox{\queue}(10,6){\thicklines
 \put(0,2){\line(1,0){4}}
 \put(0,-2){\line(1,0){4}}
 \put(4,2){\line(0,-1){4}}
 \put(6,0){\circle{3}}}

\put(0,15){\usebox\queue}
\put(5,22){$1$}
\put(8,18){\vector(1,1){8}}
\put(8,18){\vector(1,-1){8}}

\put(16,23){\usebox\queue}
\put(21,30){$2$}
\put(24,26){\vector(1,0){8}}

\put(16,7){\usebox\queue}
\put(21,14){$3$}
\put(24,10){\vector(1,0){8}}
\put(24,10){\vector(1,2){8}}                            

\put(32,23){\usebox\queue}
\put(37,30){$4$}
\put(40,26){\vector(1,-1){8}}

\put(32,7){\usebox\queue}
\put(37,14){$5$}
\put(40,10){\vector(1,1){8}}

\put(48,15){\usebox\queue}
\put(53,22){$6$}
\put(56,18){\vector(1,0){5}}

\put(12,0){(a) Network scheme}

\put(70,18)
{$
G=\left(\begin{array}{cccccc}
    \varepsilon & 0           & 0           & \varepsilon & \varepsilon 
                                                          & \varepsilon \\
    \varepsilon & \varepsilon & \varepsilon & 0           & \varepsilon 
                                                          & \varepsilon \\
    \varepsilon & \varepsilon & \varepsilon & 0           & 0 
                                                          & \varepsilon \\
    \varepsilon & \varepsilon & \varepsilon & \varepsilon & \varepsilon
                                                          & 0 \\
    \varepsilon & \varepsilon & \varepsilon & \varepsilon & \varepsilon 
                                                          & 0 \\
    \varepsilon & \varepsilon & \varepsilon & \varepsilon & \varepsilon 
                                                          & \varepsilon
  \end{array}\right)
$}
\put(80,0){(b) Support matrix}

\end{picture}

%\vspace{-2ex}
\end{center}
\caption{The fork-join queueing network model.}\label{F-FJNM}
\end{figure}

As it is easy to see, the fork-join network representation of the above 
security operation model takes the form presented in Fig.~\ref{F-FJNM}.

For the network graph, we have $ p=3 $. Therefore, we get equation 
(\ref{E-xAx}) with 
$ A(k)=(E\oplus{\cal T}_{k}\otimes G^{T}
\oplus({\cal T}_{k}\otimes G^{T})^{2})
\oplus({\cal T}_{k}\otimes G^{T})^{3})\otimes{\cal T}_{k} $.

Let us consider an arbitrary fork-join queueing network with $ n $ nodes, 
which is governed by equation (\ref{E-xAx}). We assume that the matrix 
$ G $ at (\ref{E-xAx}) has the upper triangular form. Since the network 
graph is acyclic, the network nodes can always be renumbered so that the 
matrix $ G $ become upper triangular.

Now we describe a tandem queueing system associated with the above network. We
assume the evolution of the tandem system to be governed by the equation
$$
\mbox{\boldmath $x$}(k)=B(k)\otimes\mbox{\boldmath $x$}(k-1), 
\qquad
B(k)
=
\bigoplus_{j=0}^{n}({\cal T}_{k}\otimes H^{T})^{j}\otimes{\cal T}_{k},
$$
where $ H $ is a support matrix with the elements
$$
h_{ij}=\left\{\begin{array}{ll}
        0, & \mbox{if $ i+1=j $}, \\
        \varepsilon, & \mbox{otherwise}.
       \end{array}\right.
$$

Note that both matrices $ A(k) $ and $ B(k) $ are determined by the 
common matrix $ {\cal T}_{k} $, but different support matrices $ G $ and 
$ H $. Clearly, the longest path in the graph associated with the tandem
queue is assumed to be equal $ n $.

\begin{lemma}
For all $ k=1,2,\ldots $, it holds that $ A(k)\leq B(k) $.
\end{lemma}

{\noindent\bf Proof:} 
As it is easy to verify, for any integer $ q>0 $, it holds
$$
G^{q}\leq H\oplus H^{2}\oplus\cdots\oplus H^{n}.
$$

Furthermore, since $ {\cal T}_{k} $ has only nonnegative entries on the 
diagonal, we have for any $ q>1 $,
$$
H^{q}\otimes{\cal T}_{k}\leq(H\otimes{\cal T}_{k})^{q}.
$$

By applying the above inequalities together with the condition that
$ H^{m}={\cal E} $ for all $ m>n $, we arrive at the inequality
$$
(G\otimes{\cal T}_{k})^{q}
\leq
(H\otimes{\cal T}_{k})\oplus(H\otimes{\cal T}_{k})^{2}
\oplus\cdots\oplus(H\otimes{\cal T}_{k})^{n}.
$$

Taking into account that the last inequality is valid for all $ q>0 $, we 
have
$$
{\cal T}_{k}\otimes\bigoplus_{j=0}^{p}(G\otimes{\cal T}_{k})^{j}
\leq
{\cal T}_{k}\otimes\bigoplus_{j=0}^{n}(H\otimes{\cal T}_{k})^{j}.
$$

It remains to transpose the both side of the inequality to get the desired
result.

By applying the above lemma together with the result in \cite{Krivulin2001Evaluation}, one can
prove the following statement.
\begin{lemma}
Suppose that for the acyclic fork-join queueing network, the random variables
$ \tau_{i1},\tau_{i2},\ldots $, are i.i.d. for each $ i=1,\ldots,n $ 
with finite mean $ {\rm E}[\tau_{i1}]\geq0 $ and variance 
$ {\rm D}[\tau_{i1}] $. Then the cycle time $ \gamma $ can be evaluated 
as
$$
\gamma=\max\{{\rm E}[\tau_{11}],\ldots,{\rm E}[\tau_{n1}]\}.
$$
\end{lemma}

\bibliographystyle{abbrvurl}

\bibliography{Modeling_and_performance_evaluation_of_computer_systems_security_operation}

\end{document}